\documentclass{ws-ijmpa}

\begin{document}

\markboth{S.P.~Knurenko}{Spectrum of Cosmic Rays Above $10^{17}$~eV}

\catchline{}{}{}

\title{COSMIC RAYS WITH ENERGY ABOVE $10^{17}$~eV}

\author{\footnotesize V.P.~Egorova, A.V.~Glushkov, A.A.~Ivanov, S.P.~Knurenko,
V.A.~Kolosov, A.D.~Krasilnikov, I.T.~Makarov, A.A.~Mikhailov, V.V.~Olzoyev, M.I.~Pravdin,
A.V.~Sabourov, I.Ye.~Sleptsov, G.G~Struchkov}
\address{Yu.G.Shafer Institute of Cosmophysical Research and Aeronomy 31 Lenin Ave.
\\677891 Yakutsk, Russia
\\E-mail: s.p.knurenko@ikfia.ysn.ru}

%

\maketitle


\begin{abstract}
The energy spectrum of primary cosmic rays with ultra--high energies based on the Yakutsk
EAS Array data is presented. For the largest events values of $S_{600}$ and axis
coordinates have been obtained using revised lateral distribution function. The affect of
the arrival time distribution at several axis distance on estimated density for Yakutsk
and AGASA is considered.
\end{abstract}

\section{Introduction}
Research into the cosmic rays above $10^{17}$~eV spectrum shape and into intensity in the
region of cutoff near $10^{20}$~eV predicted by Greisen~\cite{bib1}, Zatsepin and
Kuzmin~\cite{bib2} are of great importance for their sources detection. Results obtained
in various experiments~\cite{bib3,bib4,bib5} differ from each other by factor 2 and more
in absolute intensity, but their shapes are similar. At energies greater than the
GZK--cutoff results are inconsistent.  At the Yakutsk array after recent
analysis~\cite{bib3} have been carried out, there is only one event with energy estimated
to be greater than $10^{20}$~eV. To explain this contradiction with AGASA,
A.~Watson~\cite{bib6} assumed that at the Yakutsk array such showers are skipped due to
inadequate small integration time for large distances from the axis. In this work we have
studied the affect of particle arrival time distribution at different distances on
estimated density for Yakutsk and AGASA. We provided core location with adjusted LDF for
the largest events, which in average led to increase of $S_{600}$.

\section{Density Measurement At Large Distances From The Axis}
At the Yakutsk array and at AGASA a nearly similar RC--convertors are used. At the
Yakutsk array for an event to be treated, a coincedence of signals from both detectors
within 2~mcsec is required. Herewith input of convertors is closed in 2 mcsec after
coincedence. In the case when shower front is wide, this may result in underestimation of
the density. These circumstances have been pointed out by Watson in his
report~\cite{bib6}. At AGASA input is constantly open and in the case of wide signal this
may lead to density overestimation due to convertor's features.  To examine the influence
of the effects mentioned above, we have provided simulation for response of detectors at
distances $R =$ 1050, 1500 and 2000~m, based on the particle distribution approximation
obtained at AGASA~\cite{bib7}. A coefficient $K_R$ was considered --- a ratio between
density estimated with RC--convertor and the one set with program. For the Yakutsk array,
in the case of detectors with large area, registering large particle densities we have
following values: $K_{1050} = 1.05$, $K_{1500} = 0.994$, $K_{2000} = 0.76$. Same points
for AGASA: $K_{1050} = 1.065$, $K_{1500} = 1.11$, $K_{2000} = 1.2$. At 2000~m distance for
Yakutsk --- 25\% underestimation, for AGASA --- 20\% overestimation.

For the real experiment for the shower with $E_0 = 10^{20}$~eV at $R = 2000$~m about 2
particles per detector is expected. Simulation indicated, that in this case
underestimation is much less than $K_{2000} = 0.92$. It is connected with the fact that
conversion starts only after the first particle hit and at low density the effective
thickness of the shower front decreases. Probability of that the station doesn't operate
due to gap between operating of two separate detectors is more than 2~mcsec is 8.5\% and
it is lower by factor 3 than those due to Poisson fluctuations at this density. 

Simulation showed no significant underestimation of particle density for distances up to
2000~m for density measurement system at the Yakutsk array. In the case of AGASA, when
input of RC--convertor is constantly open, besides wide distribution, there is an
afterpulse contribution to density overestimation from delayng particles (probably
neutrons) together with casual additives from background muons. One can conclude from the
data in paper~\cite{bib7} that delayng neutrons can overstate the density by factor 1.37
already at 500~m and further. Background muons may cause distortions in wide range of axis
distances.  If one such particle hits within last 10~mcsec of RC--circuit discharge, then
resulting density can be overestimated by factor 2 and more independently of the real
density. The effects mentioned above are excluded at the Yakutsk array thanks to blocking
of convertor's input in 2~mcsec.

\section{Energy Spectrum Summary}
Events selection was provided as described in paper~\cite{bib3}. Showers with $\theta <
60^{\circ}$ were used. For determination of the intensity for showers with $E_0 > 4
\cdot 10^{19}$~eV, an extended area together with efficient zone outside the array was used.
It was shown~\cite{bib9}, that for showers with energy greater than $10^{19}$, LDF used
in standard procedure of axis determination badly corresponded with experimental data at
$R > 1000$~m from the axis. A modificated approximation similar to AGASA's was proposed.
In this work we provided axes coordinates determination with this adjusted LDF. As a
result, in average estimated $S_{600}$ values increased: 10\% --- for showers with axes
lying within array area and 20\% on border.

\begin{figure}
\centering
\centerline{\psfig{file=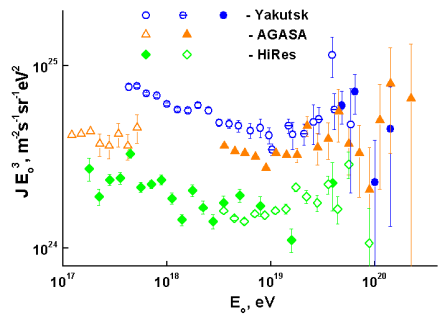,width=4.3cm,angle=90}}
\caption{Differential energy spectra according to the Yakutsk EAS array, AGASA and
HiRes.}\label{fig1}
\end{figure}

For energy estimation we used adjusted formulas of $S_{300}$ and $S_{600}$~\cite{bib10}.
On Fig.~\ref{fig1} the differential energy spectra obtained at the Yakutsk array, AGASA
and HiRes are presented. Results obtained in different experiments correspond quite well
in shape but differ in intensity. The data from the Yakutsk array near $10^{19}$~eV are
higher by factor $\simeq 2.5$ than HiRes data and $\simeq 30\%$ than AGASA's. This rather
is connected with the difference in estimation of the showers energy.
In the region of ultra--high energies the results from Yakutsk and AGASA approach. There
are 4 events registered in Yakutsk with adjusted estimated energy exceeding
$10^{19.9}$~eV. Relative errors in energy estimation in these individual showers amount
from 32\% to 46\%. If the energy is reduced by one standard error then it slightly exceeds
the $10^{20}$~eV threshold only in one event.  Therefore the relic cutoff of the spectrum
cannot be rejected based on Yakutsk EAS data.  Similar experimental errors are observed at
AGASA. According to work~\cite{bib7} their averaged value is about 20\%. Taking into account
this circumstance a conclusion was made~\cite{bib11} that yet there are too few events
recorded to approve the spectrum cutoff absense. Besides, estimations of the energy at
AGASA depend on model conclusions. The affects of observed densities mentioned above are
also not concidered yet.

\section*{Acknowledgements}
This work is supported by INTAS grant \#03--51--5112, Federal Agency of Science and
Innovations grant \#748.2003.2, RFBR grants \#02--02--16380, \#03--02--17160 and the
program \#01--30.

\end{document}